# IP Over ICN - The Better IP?

## An Unusual Take on Information-Centric Networking

Dirk Trossen[1], Martin J. Reed[2], Janne Riihijärvi[3], Michael Georgiades[4], Nikos Fotiou[5], George Xylomenos[5]

[1]InterDigital Europe, [2]University of Essex, [3]RWTH Aachen University, [4]PrimeTel, [5]Athens University of Economics & Business

*Abstract*—This paper presents a proposition for information-centric networking (ICN) that lies outside the typical trajectory of aiming for a wholesale replacement of IP as the internetworking layer of the Internet. Instead, we propose that a careful exploitation of key ICN benefits, expanding previously funded ICN efforts, will enable individual operators to improve the performance of their IP-based services along many dimensions. Alongside the main motivation for our work, we present an early strawman architecture for such an IP-over-ICN proposition, which will ultimately be implemented and trialed in a recently started H2020 research effort.

## I. INTRODUCTION

*Information-Centric Networking* (ICN) [1][2][3][4] has emerged as a novel network research area around 2006, with the promise to study alternative network architectures that are centered on information exchange rather than on endpoint-based communication, as in the current Internet.

This starting point, often called a *clean-slate approach* to re-thinking the future Internet architecture, has created a trajectory that positions ICN as a true global replacement for an IP-based Internet as we know it [5]. However, such a trajectory comes with drawbacks, many of which the ICN community has suffered from in recent years. For instance, the replacement of IP as the main internetworking protocol not only comes with the burden of heavy standardization, but also agreement among many stakeholders in the current Internet, ranging from operators and vendors over software developers and end-device makers to policymakers. Furthermore, the interconnected nature of the Internet due to its *Autonomous Systems* (AS) requires viable methods for truly scalable internetworking of individual ICN deployment islands; solutions to this important problem are currently still in their infancy. While early motivations for ICN [1] included the increasingly content-centric manner of consuming current Internet services, novel Internet services, such as those made possible by the Internet-of-Things (IoT) were also proposed as a possible driver for introducing a new internetworking layer. However, many of these promises of new services have not yet materialized.

In this paper, we consider ICN along a different trajectory of introduction into the market. Instead of aiming at a wholesale replacement of IP, akin to introducing IPv6, we propose to harness the innovation potential of IP-based applications and solutions, while benefitting from specific ICN solutions in terms of their potential for better performance compared to their IP-based counterparts. Contrary to the IP replacement goals of past ICN efforts, we focus instead on an individual operator (of a private IP network, even an enterprise one), trying to provide an answer to the question: *Is an IP-over-ICN system a better solution for IP-based services than pure IP-based networks?* Solutions that positively answer this question would primarily benefit an individual operator, not being dependent on global scale adoption, although a minimal interoperability of the final solutions would require some standardization. Not having to agree on a global change makes the answer to this question an interesting proposition for ICN.

In the following, we first shed some light on what 'better' here might stand for, outlining the space of the ICN benefits we intend to utilize. We then outline an architecture for a single operator deployment of an IP-over-ICN solution before presenting our plans for trials and test beds. Finally, we provide a brief outlook into the planned work within a newly started H2020 collaborative project in this space.

## II. 'BETTER' IN THE CONTEXT OF IP-OVER-ICN

When stating a hypothesis as outlined in our introduction, it is important to come to an understanding as to what the term 'better' entails in this context. While our work will focus more clearly on laying out formally defined *Key Performance Indicators* (KPIs) for evaluating such a hypothesis, we use the following section to provide some early insight into our understanding of key dimensions for what is meant by 'better'.

### A. Better utilisation in HTTP unicast streaming scenarios

In today's IP networks, video delivery predominantly uses HTTP-level unicast delivery from a server (or CDN sub-system) to a set of clients. This is true for both viewing of recorded videos (e.g., from YouTube) as well as live video transmissions. This unicast delivery precludes the use of more efficient multipoint delivery to a larger group, causing a growth of bandwidth in line with the growth of consumers. This linear dependence creates a pinching point for IPTV offerings, constituting a significant obstacle for their widespread deployment. By abstracting HTTP over an ICN network and utilizing the multipoint delivery capabilities of ICN [4], we will be able to significantly improve the overall utilization of the underlying transport networks and reduce operator costs. This differs from utilizing IP multicast capabilities since we preserve the HTTP unicast service abstraction, i.e., the service is realized following its original unicast semantic of delivering content from the server to a single requester.

### B. Better security and privacy for constrained applications

The use of connected devices with sensing capabilities has great potentials of improving our life, but with a cost: it creates new security and privacy threats. Consider the example of a multi-tenant building, where various sensors have been deployed. These sensors feed back towards the building management system with information related to, e.g., the building energy profile (e.g., temperature and humidity measurements), the security and safety of the building (e.g., a/v streams, motion detection, fire alarms), billing (e.g., energy consumption, number of parking slots used), and so on. Since this information is sensitive, each tenant should be able to define access control policies. Extending a constrained device to support access decisions is prohibitive, both from the performance/cost and security perspectives, as it will increase processing power requirements and energy consumption, as well as expose sensitive information (such as user credentials) to many entities. Consequently, all information has to be collected by more powerful network entities (e.g., a server) and all information access restriction mechanisms have to be implemented there, raising again security and privacy concerns. By abstracting CoAP over an ICN network, we will be able to associate security and privacy requirements with namespaces, enabling the definition of fine-grained, reusable access rules that will govern information access directly from personal gateways. Additionally, user authentication and authorization will be performed using lightweight protocols, geared towards safeguarding user privacy.

*C. Better management of virtualized network paths*

VLANs (virtual LANs) are a frequently used tool for providing dedicated 'connections' for specific services, such as for the voice, data and IPTV offerings of an operator. In a typical VLAN-based backhaul network, resources for these connections must be set up and the appropriate circuit provided to the customer. The dimensioning of these resources changes, if at all, only at longer timescales, mostly reflecting SLA changes rather than shorter-term metric fluctuations, such as resource utilization. A key reason for this lack of flexibility is that the IP-based network layer is unaware of content-level utilization. By integrating the inherent ICN resource management into the existing practice of VLAN management, we will be able to increase VLAN dynamicity. By incorporating shorter-term metrics (such as network load and congestion) and user-facing parameters (such as content and service popularity) into a single management framework, an IP-over-ICN system could improve on some KPIs, such as flexibility, service deployment time and utilization.

*D. Better (fairer) content distribution*

*Content Delivery Networks* (CDNs) are today's method of making content delivery delays acceptable to end users by placing popular content at nearby servers, using content aggregation services such as YouTube or Vimeo. This manual placement creates a barrier of entry for smaller content providers and, in particular, individuals due to the lack of exposure to the publishing APIs, which require dedicated agreements with the CDN provider and therefore a economic buying power to ensure the distribution via CDNs. Instead, smaller content providers and individuals need to aggregate their buying power through players such as YouTube,. While providing content through such aggregators might lead to improved delivery to end users, popular content might not necessarily be delivered with better quality in all cases. Furthermore, manual placement possibly wastes (caching) resources by inflexibly placing content in CDN servers, despite no local relevance or popularity for it over certain periods of time. By utilizing the explicit cache-aware resource management in ICN, as showcased in work such as [6], an IP-over-ICN solution can increase the fairness of content placement, by providing the best quality to the content that is most popular within a given resource management regime, while upholding current content placement agreements. Additionally, the scalability of caching can be improved through automating cache population, based upon learned usage, such that the most popular items are automatically placed into the cache, possibly through predictive means, even before the actual demand for them arises.

### III. A STRAWMAN ARCHITECTURE

Our efforts proposing to place IP-based services on top of an ICN-driven network build upon results from previous efforts, specifically those described in [2][4]. Hence, the reader is referred to those references for an understanding of how an ICN network would operate. In the following, we focus on how to utilize such ICN results for a system that exposes IP-based services while harnessing the benefits of ICN within its network operations.

*A. The main idea*

The driving paradigm in ICN is that *everything is information and information is everything*. When approaching the problem of providing IP-like connectivity over an ICN-enabled network, we utilize this paradigm most effectively by interpreting IP-based communication (over protocols like HTTP, CoAP, TCP or plain IP) as the exchange of information pertaining to a specific endpoint address, this endpoint address being the *name* of the information being exchanged. Let us illustrate this idea by assuming the desire of a device with IP address $A$ to send a packet to a device with address $B$. From an ICN perspective, this can be realized by interpreting the sending from A to B as a publication (of the sender) to the name $B$, while the designated receiver acts as the subscriber to the name $B$. This idea was first presented in [7]. In Section IV.B, we will outline in more detail how the aforementioned IP-over-ICN communication would be realized. For enabling any form of IP-based communication, our system architecture will support the realization of various IP-based protocols, such as HTTP, CoAP, TCP and basic IP datagram exchange, by mapping the various data structures of the underlying protocols onto suitably named objects within the ICN network.

*B. Constraints*

It is important to outline clear constraints that lead the definition of our system architecture. When positioning our efforts as one directed at a single operator that would like to improve its own IP-based service offering (along the various dimensions of 'better', as presented in Section II), we can already define an important constraint to our solution: *enable the usage of legacy IP-based devices*. In other words, our system architecture must allow for connecting existing IP-based devices to our solution. With this, we decouple our solution from the evolution of user devices, such as smartphones or desktops and avoid changes to device protocol stacks and APIs. Furthermore, we assume that *applications remain unchanged*; therefore, we do not require adaptations to application software. Lastly, in order to fully support the spectrum of IP-based services, our solution *must support any IP-based protocol abstraction*, such as HTTP, CoAP, TCP as well as sending individual IP packets. With these constraints, we focus the innovation space of improving IP-based services purely to the space of the single operator that wishes to benefit from this improvement, constituting a significant departure from previous ICN deployment goals of replacing entire IP-based infrastructure at a global scale.

However, we do see room for relaxing these constraints in selected cases. For instance, we foresee that in the upcoming area of the Internet of Things, ICN capabilities of the operator's network may reach the end device itself, fueled by recent ICN developments in this space as well as by the relatively low penetration of IoT solutions compared to IP-based solutions in the Internet. Furthermore, we can also envision the adaptation of software, particularly through browser-based plugins. This is particularly interesting for cases where interfaces would expose capabilities of an advanced ICN-driven network to such software modules, pushing data processing capabilities to the very edge of the network.

*C. System architecture*

Due to these constraints, our system architecture follows a gateway-based approach, where the first link from the user device to the network is based on IP-based protocols, such as HTTP, CoAP, TCP or IP, while the *network attachment point* (NAP) serves as the entry point into the ICN-enabled operator network, mapping the chosen protocol abstraction to ICN. We furthermore foresee a modified *ICN border GW* that establishes IP-level connectivity to peering IP networks, thus establishing true global (IP-level) connectivity for user devices connected to the operator's network. Fig. 1 illustrates this gateway-based approach and shows the main interfaces throughout the system. Driven by our constraints, the communication from an IP-enabled user equipment (UE) to the NAP takes place via standard IP-based protocols. As argued in Section III.B, we also foresee cases, such as IoT devices, where the NAP-UE communication directly utilizes ICN.

From the NAP, communication switches to ICN-based interfaces, based on the architecture outlined in [4]. These interfaces capture the interaction between the core functions of the ICN architecture, namely *Rendezvous* (RV), *Topology Management* (TM) and *Forwarding* (FN). Given the gateway-based approach, the NAP acts as the ICN client, i.e., the publisher and/or subscriber. For the former case, it utilizes the $ICN_{PR}$ interface to signal information availability to the RV, with the *name* of an information item being represented as a statistically unique label within a directed acyclic graph namespace. In the case of a positive match, the RV instructs the TM, via the $ICN_{RT}$ interface, to assemble suitable communication resources for information exchange. The TM then provides the result to the publisher via the $ICN_{TP}$ interface, which enables the NAP to publish information via the $ICN_F$ interface to the network. In the subscriber case, communication between subscriber and RV takes place via the $ICN_{SR}$ interface, although there is no explicit signaling of forwarding information (receiving suitable information is the indication for a successful subscription, unless a network error occurs, which is signaled independently).

The realization of the various interfaces depends on the various network environments that we can envision for our individual operator. For this, we utilize the concept of *dissemination strategies* [4], which allow for optimizing the core functions of the (ICN) architecture, i.e., RV, TM, and FN, to suit the particular network environment in which the information is disseminated. Of particular interest to us is a strategy that utilizes the *Software-Defined Networking* (SDN) capabilities of the underlying operator transport network, with an early version of such ICN-over-SDN mapping found in [8].

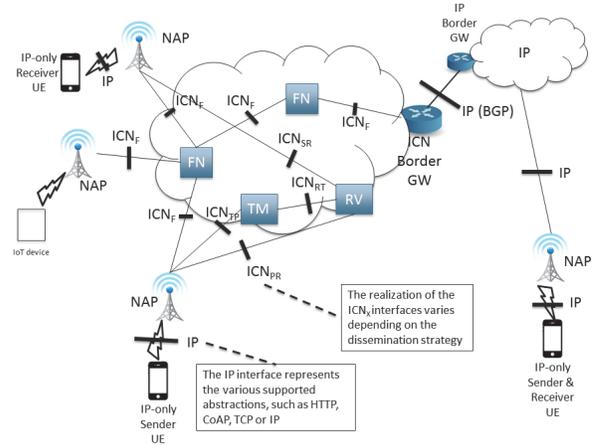

Fig. 1.  The main interfaces of our system architecture

IV. PLATFORM

The platform for validating the hypothesis presented in the introduction in realistic networks (see Section V), is based on an evolution of the Blackadder ICN platform [4], developed within the PURSUIT project, and a core software defined network using OpenFlow [9].

*A. Hitting the ground running*

The Blackadder platform has already demonstrated ICN performance at data rates up to 10 Gb/s [10]. This ICN platform follows the main system architecture in Fig. 1 through its three core functions RV, TM and FN, as shown in Fig. 2. In practice, there are multiple RV and TM instances to suit the scale of the network, while the FN function is located at every network node where packet switching is needed. The FN function is designed to be modular, in that any suitable technique can be used. Thus, a key development of our future work is to use SDN as the forwarding component of the ICN platform and to evolve the Blackadder platform to provide an ICN NAP and border GW that implements the protocol stack shown in Fig. 3. Deploying SDN allows an operator to use standard Ethernet switches, moving ICN forwarding away from a CPU intensive operation towards highly-efficient hardware switching. In the context of switching, the Blackadder platform is ideally placed to work with SDN since the TM function will work directly as an OpenFlow controller to set up *flow entries* in the SDN enabled switches. The platform will make significant use of OpenFlow v1.3 features to enable ICN supported switching and will look to OpenFlow v1.4 for providing a further evolved platform.

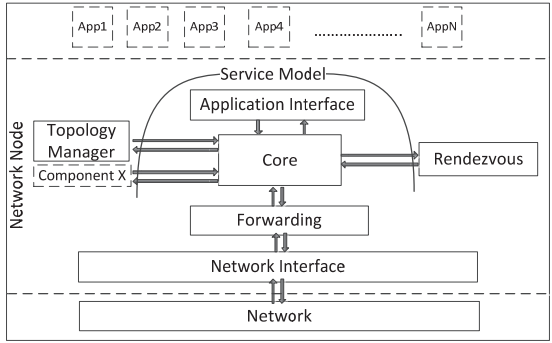

Fig. 2. The Blackadder platform as ICN core node basis

The simplified view of the protocol stack shown in Fig. 3 shows a number of abstractions for existing application network interfaces (IP, TCP, HTTP, CoAP, …). These allow existing applications to use the ICN-enabled network without changing the application interface to the network, typically a socket or HTTP level API. While we also foresee the development of native ICN applications, this approach of mapping existing application interfaces onto ICN overcomes one of the obstacles towards ICN adoption, as the end-systems and access networks do not need to be changed, as argued before. The simplest of these abstractions is the IP abstraction which is explained below.

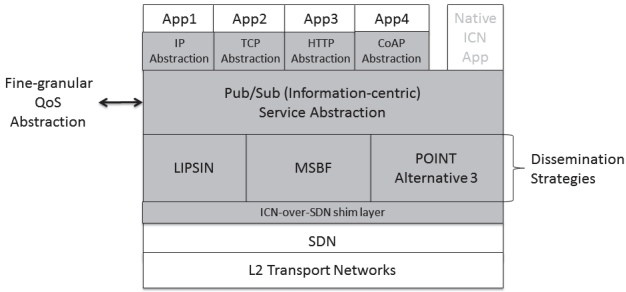

Fig. 3. Protocol stack of ICN core node

### B. Example: IP-over-ICN abstraction handler

Let us briefly elaborate on our example in Section III.A of an IP-based communication over an ICN. As the first step, it is important to outline the namespace over which the ICN network will exchange IP-based information. Fig. 4 illustrates this namespace, following the conventions in [4] and defining its own root. The namespace is divided into communication with IP devices connected to the ICN network (represented through the *I* node) and outside the ICN network (the *O* node). The subnetwork structure (here of IPv4 addresses) is represented as *scopes* of information [4], allowing for subscribing to communication with entire subnets at once through the capabilities of the ICN architecture.

The namespace is utilized by two elements in the system architecture, namely the NAP and the ICN border GW. As noted before, the communication between the IP-enabled device and the NAP is purely IP-based, i.e., DHCP and other mechanisms are utilized for IP address space allocation. For each IP address that the NAP locally assigns, it will act as the subscriber towards the ICN network, ready to receive any information being sent to the IP address. For this, it will determine the appropriate label according to our namespace and subscribe to it. With this, any IP packet being sent to an IP address allocated to a locally attached IP device will arrive at the NAP serving it. Furthermore, the ICN border GW will subscribe to the *O* scope of the namespace, which will result in receiving any IP packet sent from a locally connected IP device to an IP address outside of the operator's domain.

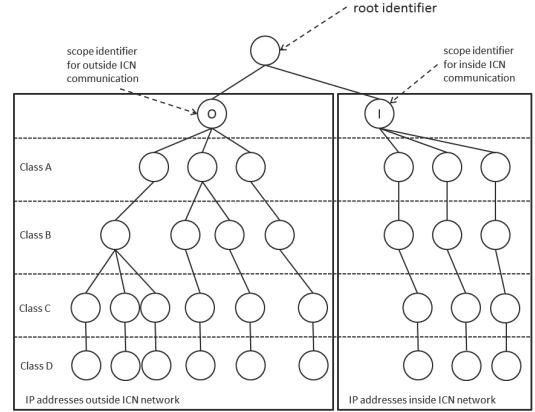

Fig. 4. IP-over-ICN namespace

For sending IP packets, we distinguish two cases, namely that of an *internal* or *external sender*. In the former case, the packet will be sent from an IP-enabled device to its local NAP. The NAP will determine the suitable ICN name for the destination address of the IP packet it received (utilizing the namespace of Fig. 4, i.e., either placing it under the *I* or *O* scope of the namespace). It will then encapsulate the IP packet in the payload of the ICN packet and publish the latter to the determined name. In the case of an internal receiver, the packet will arrive at the NAP that has previously subscribed to the appropriate IP address name (on behalf of its locally attached IP device), which in turn will decapsulate the IP packet and relay it over its local link. If the packet was destined to an ICN-external IP device, it will arrive at the ICN border GW (due to the latter subscribing to the *O* scope of the namespace), which in turn will decapsulate the IP packet and relay it over the appropriate peering link. In the case of an external sender, the packet will arrive at the ICN border GW, which in turn will encapsulate the packet as an ICN payload, determine the appropriate ICN name (under the *I* scope of the namespace) and publish it towards the ICN network.

## V. TESTBEDS AND EVALUATION

As discussed in Section II, our key objective is to critically evaluate our hypothesis that running IP-over-ICN results in a 'better' networking experience for the major stakeholders, compared to the present-day TCP/IP networking with classical routing and switching. Our evaluation work has two distinct aspects to it, *validation* and *quantitative evaluation*. Validation consists of verifying that the entire system satisfies all the functional requirements imposed on it. These include the capability of end users to employ unmodified UEs to access the various services, many of the requirements on improved privacy and security, and so on. In addition to local deployments, many of these will be validated using our extensive overlay testbed infrastructure. This testbed is constructed using VPN connections between 10 different sites, each equipped with several physical and virtual machines on which our platform can be deployed. This overlay testbed will play a major role in our development efforts, facilitating rapid integration and prototyping for the platform engineering work.

For quantitative evaluation, a different approach is needed, measuring the performance of our system against a carefully selected collection of KPIs measuring both end user quality of experience, as well as different performance aspects of interest to the network operator, all for diverse applications. Examples of the former include raw network performance measures such as throughput, latency, and jitter, but also more perceptual measures such as time to render for interactive browsing. For the latter group of KPIs, we will include aggregate performance metrics such as overall network capacity and percentage of signalling overhead, but also techno-economic aspects such as estimates of the network cost, predictability of traffic loads, and traffic engineering efforts needed. In order to obtain high quality data on these KPIs, a dedicated testbed infrastructure is needed, the performance of which is not influenced in an uncontrolled manner by the usage of the underlying network, as is the case for our overlay testbed.

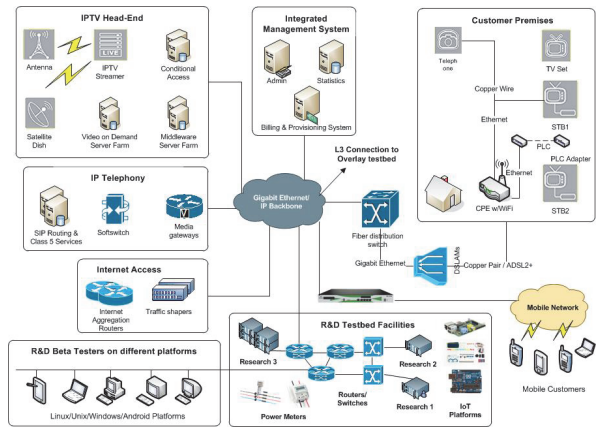

Fig. 5. Network operator's testbed facilities

For these reasons, the quantitative evaluation phase is planned to be conducted by deploying and running test trials of the system in an actual production network of a network operator aiming towards large scale validation involving real users. Fig. 5 shows how the R&D testbed facilities at PrimeTel PLC are connecting the real network through an IP backbone, from where it could also connect to other servers of interest, e.g. for IPTV, VoD, Content Services etc. The R&D testbed has a number of servers for test purposes co-located at one of the company's main datacenters, offering numerous connectivity options through typical routers, switches and other typical network equipment. Moreover, R&D engineers will, together with external users, act as beta-testers to access the R&D testbed facilities via a number of different platforms. To make the trial multi-domain, PrimeTel may further connect to the overlay testbed through L3 connectivity to better illustrate IP-over-ICN use case scenarios of interest, such as ones related to unicast streaming, multipath streaming etc. In such an integrated testbed both, content servers or content headless clients, could be used at various remote locations at the domains of other testbed partners for further testing.

VI. CONCLUSIONS

Although implementing IP services over ICN is not a novel proposition, its realization over a carrier-grade network is both novel and groundbreaking, since it places the deployment of ICN on a new trajectory, away from the wholesale replacement of IP across the entire Internet. While in principle a simple idea, its ramification can be significant in terms of benefits to a single operator's service offering. In this paper, we outlined a few of these expected benefits, presented a strawman architecture that would realize this proposition and exemplified the operation of IP-based services over an ICN infrastructure, which we intend to realize in the trial and testbed activities that we presented. The efforts towards proving or negating the hypothesis will realize the various components shown in this paper within a newly established H2020 funding effort.

ACKNOWLEDGMENTS

Many of the ideas presented in this paper stem from discussions among POINT consortium partners. Thanks goes to all contributing to these discussions, eventually leading to POINT being funded through EC H2020 ICT project 643990.